\documentclass[12pt]{iopart}

\usepackage{iopams}  
\newcommand{\be}{\begin{equation}}
\newcommand{\ee}{\end{equation}}
\newcommand{\beqn}{\begin{equation}}
\newcommand{\eeqn}{\end{equation}}
\newcommand{\bea}{\begin{eqnarray}}
\newcommand{\eea}{\end{eqnarray}}

\newcommand{\Ohat}{O}

\begin{document}

\title[`Non-observables' for nuclei]{How should one formulate,
extract, and interpret `non-observables' for nuclei?}

\author{R.J.\ Furnstahl$^{1}$ and A.\ Schwenk$^{2,3,4}$}

\address{$^{1}$ Department of Physics, The Ohio State University, 
         Columbus, OH 43210, USA}
\address{$^{2}$ TRIUMF, 4004 Wesbrook Mall, Vancouver, BC, V6T 2A3, Canada}
\address{$^{3}$ ExtreMe Matter Institute EMMI, GSI,
         64291 Darmstadt, Germany}
\address{$^{4}$ Institut f\"ur Kernphysik, TU Darmstadt,
         64289 Darmstadt, Germany}

\ead{furnstahl.1@osu.edu, schwenk@triumf.ca}

\begin{abstract}
Nuclear observables such as binding energies and cross sections can
be directly measured.  Other physically useful quantities, such as
spectroscopic factors, are related to measured quantities by a
convolution whose decomposition is not unique.  Can a framework
for these nuclear structure
`non-observables' be formulated systematically so that
they can be extracted from experiment with known uncertainties and
calculated with consistent theory?  Parton distribution functions in
hadrons serve as an illustrative example of how this can be done.  A
systematic framework is also needed to address questions of
interpretation, such as whether short-range correlations are
important for nuclear structure.
\end{abstract}

\section{Nuclear observables and non-observables}

In quantum mechanics, an observable is typically defined as a physical
property that can be measured.  The usual examples include energy,
momentum, and angular momentum.  But what about familiar nuclear
quantities such as momentum distributions and spectroscopic factors?
Although one often says that momentum distributions are ``measured'',
in fact they must be extracted from data.  The general structure is
that a measured quantity such as a cross section is decomposed as a
convolution of subsidiary pieces, usually based on a factorization
principle.  This decomposition is not unique, and so we refer here to
the extracted quantities as `non-observables'.  The quotes are
intended to soften the implication that it is improper to talk about
them; nevertheless, unless the conventions (e.g., scale and scheme
dependence) are controlled and specified, there will be ambiguities
that will be entangled with the structure and reaction approximations.
The challenge is to formulate and carry out experimental extractions
and theoretical calculations of non-observables systematically and
consistently.

The theoretical ambiguities for non-observables that take the form of
a matrix element of an operator $\Ohat$ (e.g., the momentum
distribution $\Ohat(k) = a^\dagger_k a_k$) are manifested by
considering unitary transformations.  We are free to apply a
(short-range) unitary transformation $U$ to $\Ohat$ and to states
$|\Psi_n\rangle$ with the result that matrix elements are invariant:
\be
O_{mn} \equiv \langle \Psi_m | \Ohat | \Psi_n \rangle
       = \bigl( \langle \Psi_m | U^\dagger \bigr) \,
          U \Ohat U^\dagger \, \bigl( U | \Psi_n \rangle \bigr)
       = \langle \widetilde \Psi_m | \widetilde O | \widetilde \Psi_n
         \rangle \,.
\ee
But the matrix elements of $\Ohat$ itself between the transformed
states are in general modified (unless $O$ is a conserved charge):
\be
  \widetilde O_{mn} \equiv 
  \langle \widetilde\Psi_m | O | \widetilde \Psi_n \rangle
  \neq O_{mn} \,.
\ee
In a low-energy effective theory, there is no preferred set of states
(equivalently, there is no preferred Hamiltonian) so transformations
that modify short-range unresolved physics generate equally acceptable
states. This is for example discussed in the context of nuclear forces
in Ref.~\cite{Bogner:2009bt}. Thus to the extent that differences
between $O_{mn}$ and $\widetilde O_{mn}$ are important (which will
depend on the process and kinematics), one cannot unambiguously refer
to one as a measured quantity without specifying the basis states used
and the corresponding Hamiltonian (including regularization and
renormalization scheme).%
\footnote{The field-theoretic counterpart to this argument
in terms of field redefinitions is presented in the context of 
momentum distributions and occupation numbers in 
Refs.~\cite{Furnstahl:2000we,Furnstahl:2001xq}.}
The same considerations apply to any sort of wave function overlap.
A framework to specify the relevant conventions and relate
different choices is well developed for parton distribution functions 
but its counterparts do not
yet exist for nuclear non-observables.
The challenge is to formulate such frameworks.

These considerations are critically important for non-observables such
as nuclear spectroscopic factors, which are extracted from
experimental measurements and used to compare data to data and data to
theory.  In particular cases the extracted values may vary from
essentially convention independence to highly scheme dependent;
precision analysis of nuclear data requires a rigorous framework to
assess these uncertainties and allow robust comparisons.  Interpreting
the physics implications of non-observables also requires knowing how
they evolve with transformations of nuclear Hamiltonians that change
the resolution.

\section{Nuclear and other examples}

In nuclear physics, there is a wealth of physics involved in analyzing
spectroscopic factors.  Examples include comparing the ratio of
measured to calculated knock-out cross sections as a function of
isospin; focusing on long-range characterizations of states, such as
the Hoyle state as a triple-alpha cluster state (in a different basis
this state may look very complicated); or using spectroscopic factors
as a concept to gain understanding of the nature of single-particle
dominated versus more complex states in nuclei.  For a proper
comparison and rigorous interpretation of spectroscopic factors it is
important to have control over their extraction and calculation.

We can gain insight into the issues with spectroscopic factors by
considering the simplest nucleus, the deuteron.  The D-state
probability of the deuteron is a non-observable; it depends on the
short-range tensor strength that changes under unitary transformations
of nuclear forces and the associated transformations of basis
states. The D-state probability is a spectroscopic factor for the
D-state part of the wave function, so this is an example of a general
model dependence of spectroscopic factors.  The deuteron example, with
the resolution dependence of the S- and D-state probabilities (shown,
e.g., in Fig.~57 of Ref.~\cite{Bogner:2009bt}), demonstrates that there
may be a large theoretical uncertainty (roughly a factor of two for
the D-state deuteron probability) associated with spectroscopic
factors that are sensitive to short-range parts, and a significant
theoretical uncertainty (of order 10\% for the S-state deuteron
probability) for spectroscopic factors that probe mainly the
long-range parts of nuclear forces.

In addition, spectroscopic factors rely on a convention for the
long-range parts in nuclear Hamiltonians, such as pion exchanges or a
pionless theory. Both describe very low-energy scattering with similar
accuracy, but the corresponding long-range parts of wave functions
will differ.  We stress the importance of separating midrange physics
from short-range contributions in the discussion of spectroscopic
factors. For example,
midrange will receive contributions from one- and multi-pion
physics in chiral effective field theory (EFT).

In systems with a large separation of scale, spectroscopic factors are
effectively measurable (that is, there are only small systematic
uncertainties) because the unitary transformations that change
spectroscopic factors lead to shifts of order $(k R)^n$, with $k R \ll
1$ (where $k$ is a typical momentum scale and $R$ denotes the range of
interactions). An excellent example is a precision measurement of the
closed-channel fraction of cold atom pairs across a broad Feshbach
resonance, obtained from radio-frequency spectroscopy where initial
and final state interactions are
negligible~\cite{PhysRevLett.95.020404}.  Similarly, the momentum
distribution for strongly interacting cold atom gases near the
unitarity limit follows a universal power law $1/k^4$ as $k
\rightarrow \infty$ as long as $k R \ll 1$.  
As for cold atoms near unitarity, when there
is a large separation between the energy scales of interactions and
the excitation of the composite particles, the
impulse approximation will be good and the separation of the momentum
distribution from experimental data will also be clean.  But this is
not the case for nuclei, which therefore require greater care.

\section{Parton distribution functions as a paradigm}

We propose the framework of parton distribution functions (PDF's) as
a paradigm for nuclear non-observables.  The general scenario is that
experimental cross sections are expressed as a well-defined
convolution.  The PDF analysis is based on the expectation that part
of the convolution can be calculated reliably for given experimental
conditions so that the remaining part can be extracted as a universal
quantity, which can then be related to other processes and kinematic
conditions.  In the case of hard-scattering processes with a large
momentum transfer scale $Q$, factorization allows a separation of the
momentum and distance scales in the reaction.  (In short, the time
scale for binding interactions in the rest frame are time dilated in
the center-of-mass frame, so the interaction of an electron with a
hadron in deep-inelastic scattering is with single non-interacting
partons.)  The short-distance part can be calculated systematically in
low-order perturbative QCD and the long-distance part identified as
PDF's, which are basically momentum distributions for partons (i.e.,
quarks and gluons) in hadrons.

The PDF's are typically extracted from global fits to experimental
data.  Related quantities to PDF's are fragmentation functions, parton
distribution amplitudes, generalized parton densities, which all
summarize the universal non-perturbative parts of the physics.  Since
they are universal, the same PDF's appear in all reactions, so once
determined they can relate processes such as the deep inelastic
scattering of leptons, Drell-Yan, jet production, and more.  Thus one
can measure them in a limited set of reactions and then perturbative
calculations of hard scattering and PDF evolution enable first
principles predictions of cross sections for other
processes~\cite{Collins:2003fm}.

The rigorous framework developed for PDF's and related functions offers
important lessons for analogous low-energy nuclear quantities.  The
momentum distribution for a given hadron is not unique: there is
dependence on $Q^2$, which serves as the resolution scale and can be
changed by renormalization group (RG) evolution, and the PDF analysis
at next-to-leading order must be performed in a specific
renormalization and factorization scheme (typical choices are MS and
DIS)~\cite{Campbell:2006wx}.  To maintain consistency, any
hard-scattering cross section calculations that are used for the input
processes or that use the extracted PDF's have to be implemented with
the \emph{same} scheme.  There is careful treatment of the
uncertainties in the PDF's.  It is \emph{not} considered sufficient to
just compare different extractions.  Instead, the Lagrange Multiplier
and Hessian techniques have been developed to estimate PDF
uncertainties (see Ref.~\cite{Campbell:2006wx} for brief explanations
and original references).  Finally, in expressing a PDF in terms of a
quark correlation function there are non-trivial details that must be
done correctly; as emphasized by Collins~\cite{Collins:2003fm}, naive
factorization is not adequate.

Each of these features has an analog in calculating spectroscopic
factors and other nuclear non-observables.  Can we formulate our
theory to have the same control as with PDF's using factorization?  An
underlying question is whether the necessary ingredients
(corresponding to asymptotic freedom, infra-red safety, and
factorization for PDF's, see Ref.~\cite{Tung:2001cv}) are present.  
If not,
there will be intrinsic limitations that have to be quantified.  The
seeds for such a framework have been under development for some years.
These include EFT methods to consistently calculate
structure and operators~\cite{Epelbaum:2008ga} and 
RG methods that can change the resolution~\cite{Bogner:2009bt}.
Extending these methods to the problems discussed here is an important
open task for nuclear theory.

\section{Interpreting non-observables for nuclei}

A rigorous framework for extracting nuclear non-observables is needed to
offer clear interpretations.  A case in point are recent experiments
that are interpreted as providing definitive evidence for the effects
of short-range correlations in nuclei.  This interpretation raises many
questions, as these correlations are features of the nuclear wave
function at short distances, which are particularly dependent on the
choice of the Hamiltonian (e.g., the resolution).  How is this physics
reconciled with approaches using chiral EFT or low-momentum
interactions, for which short-range correlations are greatly
suppressed, or with ``mean-field'' energy-density functionals?  For
low-momentum interactions, the unitary transformations leave
observable cross sections unchanged by construction. Does one then
simply have different interpretations at different resolutions 
(e.g., simple operator and complicated wave function versus 
complicated operator and simple wave function)
or are
there basic limitations to what can be concluded?  Is the extraction
from experiment better controlled at certain resolutions? Does
factorization work better for a range of resolutions in nuclear
Hamiltonians?  The challenge is to develop a systematic framework that
can be applied at different resolutions, to address these questions
and enable the theoretical ambiguities involved in non-observables to
be quantified.

While implications from high-momentum components in nuclear wave
functions have been claimed for nuclear structure at both normal and
higher densities (e.g., neutron stars), one should be cautious about
attributing too much to resolution-dependent quantities.  After all,
with parton distributions one would not talk about the results at a
particular $Q^2$ as being ``the'' quark or gluon momentum distribution
also for lower or higher $Q^2$.  In addition, we note the importance
of distinguishing short-range versus long-range (or mid-range)
correlations.  The issue is whether one understands the physics at the
corresponding resolution scale, which makes it possible to attribute
the result to the correct physics.  This is relevant for recent
experiments at Jefferson Lab that measure the differences in
correlations between proton-proton and proton-neutron pairs by looking
at the ratio of differential cross sections
$d\sigma(e,e'pn)/d\sigma(e,e'pp)$, which shows a pronounced preference
for $pn$ pairs over $pp$ pairs~\cite{Subedi:2008zz}.  
It is important to disentangle what
is due to effects of long-range (low-momentum) pion-exchange tensor
forces as opposed to the high-momentum reaction physics or short-range
effects usually associated with a strong short-range repulsion in
nucleon-nucleon (NN) interactions.

It is sometimes said (e.g., see Ref.~\cite{Frankfurt:2009vv}) that
short-range correlations are ``hidden'' in the parameters of
low-energy effective theories (e.g., an EFT).  Presumably ``hidden''
is in the sense of integrated out; that is, contributions from loop
momenta are shifted into coupling constants as the resolution is
decreased.  When is it necessary (or at least desirable) to ``unhide''
this physics when doing low-energy nuclear physics?  In particular,
is it relevant
when calculating low-energy nuclear structure, such as binding
energies and low-lying excitations?  A systematic framework is needed to
disentangle what is hidden but known from what is 
unconstrained short-distance physics.

In interpreting occupation numbers or momentum distributions extracted
from experiment, a comparison is often made to independent-particle
models, where occupation numbers are either zero or one.  
This in turn sometimes leads to criticism of mean-field
energy-density functionals (EDF) because of the apparent contradiction
of fractional occupation numbers extracted and those in the EDF being
equal to zero or one (ignoring
pairing here).  However, the theoretical underpinning of EDF
approaches is Kohn-Sham density functional theory (DFT), for which
(without pairing) the occupation numbers \emph{are} zero or one,
regardless of the degree of correlation.  At the same time, there are
issues with using Kohn-Sham single-particle wave functions for
non-Kohn-Sham observables (although calculations of single-particle
levels in DFT have shown promise, even if not fully justified by the
theory).  How can we resolve these conflicting views?  A key
development would be a unified framework for DFT and experiments such
as $(e,e'p)$.  This would also help to distentangle the role of
short-range versus long-range correlations in the theoretical
calculation of occupation numbers.  Recent results for the quenching
of spectroscopic factors suggest that long-range correlations may be
more dominant than previously realized~\cite{Barbieri:2009ej}.
Being able to make robust comparisons at different resolutions
will be essential in addressing these issues.
 
A final example of a non-observable that requires careful treatment is
the extraction of NN potentials from recent lattice QCD calculations.
In this case it is computational rather than experimental data that is
analyzed.  We first note that the short-range NN interaction is a
non-observable, which is immediately evident from the experience with
unitary transformations~\cite{Bogner:2009bt}.  In the case of the
lattice calculations, a choice has to be made of quark fields that are
to be used as interpolating operator for the nucleon (which is chosen
to have a non-zero overlap with a nucleon state that will dominate at
large Euclidean times).  Different choices will lead to different
potentials so one must be cautious in drawing conclusions.
Nevertheless, this could be a powerful tool to gain physical insight,
in particular given the many knobs to change QCD parameters on the
lattice.

\section{Final comments}

The systematic formulation of `non-observables' such as spectroscopic
factors and related quantities is an important open problem for
nuclear structure theorists.  The analogy to parton distribution
functions seems to us to be a promising avenue to pursue.  However,
the development and application
of corresponding frameworks that can address questions
such as the validity of factorization approximations with different
Hamiltonian resolutions will require progress on
other open questions of nuclear structure outlined in this volume.  
Finally, we
note that resolution dependence for a
given non-observable does not mean that a particular choice
of conventions cannot be advantageous, just as in field theory contexts
the choice of a particular gauge may be most illuminating or
predictive.  But having a rigorous framework will enable this choice to be made
in a controlled way.

\section*{Acknowledgments}

This work was supported in part by the National Science Foundation
under Grant No.~PHY--0653312, the UNEDF SciDAC Collaboration under
DOE Grant DE-FC02-07ER41457, the Natural Sciences and Engineering
Research Council of Canada (NSERC), and by the Helmholtz Alliance
Program of the Helmholtz Association, contract HA216/EMMI ``Extremes
of Density and Temperature: Cosmic Matter in the Laboratory''.  TRIUMF
receives federal funding via a contribution agreement through the
National Research Council of Canada.

\section*{References}

\bibliographystyle{unsrt}

\bibliography{vlowk_refs}

\end{document}